\documentclass[review]{elsarticle}
\usepackage{amsmath}

\usepackage{lineno,hyperref}
\modulolinenumbers[5]

\usepackage[nodots]{numcompress}
\usepackage{amsfonts}
\usepackage{mathrsfs}
\usepackage{latexsym,amssymb}
\usepackage{graphicx}
\usepackage{subfigure}
\usepackage{array}
\usepackage{float}
\usepackage{multirow}
\usepackage{tabularx}
\usepackage{booktabs}
\biboptions{sort&compress}

\providecommand{\U}[1]{\protect\rule{.1in}{.1in}}

\journal{Applied Mathematics and Computation}

\begin{document}

\begin{frontmatter}

%\title{Fault tolerant control based on adaptive neural observer for a 3-DOF helicopter}
%
%
%%\author{Yujia~Wang, Tong~Wang, Xuebo~Yang, and~Jiae~Yang
%%  \thanks{The authors are with the Research Institute of Intelligent Control and Systems, Harbin Institute of Technology, Harbin, China.}}
%\author[mymainaddress]{Xuebo Yang}
%\ead{xueboyang@hit.edu.cn}
%
%\author[mymainaddress]{Yujia Wang\corref{mycorrespondingauthor}}
%\ead{WangYujia_05@163.com}
%
%\author[mymainaddress]{Tong~Wang}
%\ead{twang@hit.edu.cn}
%
%\author[mymainaddress]{Jiae Yang}
%\ead{yangjiae1111@163.com}
%
%\address[mymainaddress]{School of Astronautics, Research Institute of Intelligent Control and Systems,Harbin Institute of Technology, Harbin 150001, China}

\title{A Novel Method to Design Controller Parameters by Using Uniform Design Algorithm}

% Group authors per affiliation:

%\author{Yujia~Wang, Tong~Wang, \emph{Member, IEEE},  Xinghu~Yu, Xuebo~Yang, \emph{Member, IEEE}
%\thanks{Yujia Wang, Tong Wang and Xuebo Yang are with the Research Institute of Intelligent Control and Systems, Harbin Institute of Technology, Harbin, China. (e-mail: twang@hit.edu.cn)}}

\author[mymainaddress]{Yujia Wang\corref{mycorrespondingauthor}}
\cortext[mycorrespondingauthor]{Corresponding author}
\ead{iiwgwsw@163.com}

\author[mymainaddress]{Tong Wang}
\ead{twang@hit.edu.cn}

\author[mymainaddress]{Jiae Yang}
\ead{yangjiae1111@163.com}

\author[mymainaddress]{Xuebo Yang}
\ead{xueboyang@hit.edu.cn}

\address[mymainaddress]{School of Astronautics, Research Institute of Intelligent Control and Systems,Harbin Institute of Technology, Harbin 150001, China}

\begin{abstract}
 Parameter selection is one of the most important parts for nearly all the control strategies. Traditionally, controller parameters are chosen by utilizing trial and error, which is always tedious and time consuming. Moreover, such method is highly dependent on the experience of researchers, which means that it is hard to be popularized. In this light, this paper proposes a novel parameter searching approach by utilizing uniform design (UD) algorithm. By which the satisfactory controller parameters under a performance index could be selected. In this end, two simulation examples are conducted to verify the effectiveness of proposed scheme. Simulation results show that this novel approach, as compared to other intelligent tuning algorithms, excels in efficiency and time saving.
\end{abstract}

\begin{keyword}
 Parameter section; Uniform design algorithm; Nonlinear systems; Control design.
\end{keyword}

\end{frontmatter}

%\linenumbers

\section{Introduction}
%For many years, unceasing efforts \cite{Visioli2001Tuning,vsitum2014optimization,patrascu2017self}  have been made to find effective control strategies that are able to ensure  fast convergence of the system and stability in finite time. However, the time for the controller to stabilize the system is much shorter than the time spent in choosing the controller parameters and thus, the shortening of control time seems to be of limited significance.

Up to now, a large variety of control strategies \cite{8613854,6269080,8571243,5410142,8901393} have been proposed to meet different control objectives. Besides the control methods, the control parameters have significant influence on the control performance. When they are not designed appropriately, the system performance would decrease or even diverge. Accordingly, it is of vital importance to find an efficient way to search the appropriate controller parameters in the field of control. Typically, operators adjust controller parameters in accordance with their experience, which often leads to the situation that there is no unified optimal result and a considerable amount of time and energy needed to be taken. At the same time, when the model of the controlled system is nonlinear, strongly coupled and multivariable, the control parameter design would be more difficult.

%the fact that various existing control methods and the responding satisfactory control parameters are different increases the difficulty of tuning control gains. However, when the model of the controlled system is nonlinear, strongly coupled and multivariable.    applied in many control systems, such as spacecraft \cite{Yang2012An,jia2016optimization}, servo hydraulic system \cite{vsitum2014optimization} and others \cite{herrero2002optimal,patrascu2017self}. However, it is not a perfect strategy for its disadvantages of local optimum and premature convergence, especially when the crossover and mutation probabilities are selected inappropriately. Furthermore, GA would lead to enormous computational loads \cite{kawabe1997real,krohling1997designing}, as do other optimization algorithms such as particle swarm optimization (PSO) and its improved algorithms \cite{ling2007particle,ko2008pso,gaing2004particle}, bio-inspired stochastic search algorithm \cite{merheb2012novel,neumann2013bioinspired,merheb2013passive}, etc. It is

Ziegler Nichols as a classical method is first introduced to tune PID parameters. However, it is quite hard to work out optimal values in practical engineering and it cannot be applied to tune controller parameters of other control strategies. In this connection, researchers began to explore intelligent methods, and a large number of techniques have been evolved. Genetic Algorithm (GA) is a common technique to intelligently adjust control parameters. By selecting the appropriate fitness function, desired control parameters can be obtained through intelligent search so as to achieve relatively short settling time, overshoot and zero steady state error. Such intelligent search methods have been
applied in many plants, such as spacecraft, servo hydraulic system, etc. However, it is not a perfect strategy for its disadvantages of local optimum and premature convergence, especially when the crossover and mutation probabilities are selected inappropriately. Furthermore, GA would lead to enormous computational loads \cite{8540853,Jia2016Optimization}, as do other optimization algorithms such as particle swarm optimization (PSO) and its improved algorithms \cite{8377479,Tang2011Parameters}, bio-inspired stochastic search algorithm \cite{7932143}, etc. It is
undeniable that the existing intelligent algorithms are instrumental in helping people to tune controller parameters, but they often fall into local optimum and involve a great deal of computation to find the best solution especially when there are many designed control parameters.

To overcome the challenges mentioned above,  a novel research technique by using UD is proposed in this paper. UD as a method to solve the problem of designing experiments was proposed by Fang and Wang in China in the late 1970s \cite{fang2001orthogonal,wang2005note,fang2000uniform,fang1980uniform}. Consequently, this number theory-based method has been widely used in the existing results \cite{liang2001uniform,Wang2007The,Wang2018Investigating}. It differs from other experimental design method in that it only takes into consideration the uniform distribution of test points. Compared with orthogonal design and overall design method, it is enable to largely reduce the experiment times and reflect the relationship between the factors and the experiment indicators, simultaneously. The research results suggest that the factorial, orthogonal and UD have test-sample-averaged errors of $0.002\%$, $0.029\%$ and $0.023\%$ respectively in the case of few experimental factors and levels. As sample factors and levels increase, the factorial becomes unacceptable due to excessive test points, but the test-sample-averaged error of orthogonal and UD are $0.42\%$ and $0.15\%$, respectively. Thus, UD is regarded as a quite effective way to find a satisfactory combination among large amounts of samples.
% \cite{ju2010artificial}
%@article{ju2010artificial,
%  title={Artificial neural network model of airfoil aerodynamic performance using design of experiments},
%  author={Ju, YP and Zhang, CH},
%  journal={Acta Aeronaut Astronaut Sin},
%  volume={31},
%  number={5},
%  pages={893--898},
%  year={2010},
%  month={May.}
%}

Inspired by the above mentioned methods, a controller parameter research method by using UD algorithm is proposed. The contributions of this work are given as follows:
\begin{itemize}
\item [1)] The proposed method not only can be used to design the parameters of PID method, but also can be used to design the parameters of any other control strategies.
\item [2)] Compared with the existing intelligent search methods, the proposed method can avoid local optimum. Also, it is more time-saving and easier to use.
\end{itemize}

The structure of this paper is organized as follows. Problem formulations and some preliminaries about UD algorithm are demonstrated in section 2. Section 3 details how to apply UD algorithm to obtain the optimal controller gains. Section 4 gives the simulation results to verify the merits of the proposed method, followed by some conclusions in section 5.

%%%%%%%%%%%%%%%%%%%%%%%%%%%%%%%%%%%%%%%%%%%%%%%%%%%%%%%%%%%%%%%%%%%%%%
\section{Problem Formulations and  Preliminaries}

%\begin{figure}[b]
%\centering
%\includegraphics[width=7.5cm,height=4cm]{controlProcess.eps}\caption{ Block diagram of control system \label{fig1}}%%
%\end{figure}

The controller parameter design is very critical, but it is not an easy task especially when the model of the controlled system is nonlinear, strongly coupled and multivariable. The object of this paper is to present a novel method by using UD algorithm to search the controller parameters under a specific performance index.

Suppose each control parameter $k_i$, $i=1,2,\cdots,s$ can be selected from a set $A_i=\{a_{i1},a_{i2},\cdots,a_{ij}\}$, $j=1,2,\cdots,q$. Then, the number of all the combination of control parameters is $n=q^s$, which can be extremely large when the number of control parameters $s$ and and the number of elements in optional set $q$ are too large. In such case, it would waste a large amount of time to find a combination such that the satisfactory control performance is achieved. However, UD algorithm can select representative points from a large quantity of data within a acceptable error. It should be noted that when selecting representative points, neat comparability and uniformly dispersed are two important evaluation criteria. The former denotes the representativeness of these test points, and the latter facilitates the analysis of data. To reduce the number of combinations, the UD algorithm only retains the characteristic of uniform dispersion which enables the test points to be evenly distributed in the test range. The UD algorithm mainly includes two parts, with the first one being how to design uniform design table, and the second one being how to create the use table of uniform design table.

\subsection{Uniform design table}
Generally, the uniform design table is remarked as ${U_n}({q^m})$, where $n$ is the number of rows of a table, which refers to experiment time (the number of combination of control parameters), the value $m$ represents the total number of columns (factors) (the number of control parameters, i.e., when PID algorithm is used as control method for the system $\dot x=f(x,u)$, $x\in R$, then, the number of control parameters $m=3$. That is, proportional gain, integral gain and differential gain.), $q$ denotes the level number of each factor (the number of elements in each optional set). Usually, the level number $q$ is equal to the number of test times $n$ and thus, the uniform design table can also be written as ${U_n}({n^m})$.

%Generally, the uniform design table is remarked as ${U_n}({q^m})$, where $n$ is the number of rows of the table, the value $m$ represents the number of columns of the table, $q$ denotes the level number of each factor (the number of elements in each optional set). Usually, the level number $q$ is equal to the number of test times $n$ and thus, the uniform design table can also be written as ${U_n}({n^m})$.

Each uniform design table can be regarded as a matrix with $n$ rows and  $m$ columns. Each column is a non-repeatable combination of $\left\{ {1,2, \cdot \; \cdot \; \cdot ,n} \right\}$, each row is the subset of $\left\{ {1,2, \cdot \; \cdot \; \cdot ,n} \right\}$. It should be noticed that this subset may not be an appropriate subset for the reason that there are many combinations that meet such requirement. Thus, it is very critical to find an effective way to determine which combination is the best one. For the sake of reducing computation and constructing a better uniform design table, Fang Kai Tai and Wang Yuan decided to use the GLP (Good Lattice Point) method after repeated theoretical proofs and experiments \cite{fang2001orthogonal,wang2005note,fang2000uniform,fang1980uniform}. The specific steps to design uniform design table are described as follows:

\begin{enumerate}
\item[(i)] Give
%$m$ factors, where $m$ is a positive integer. The level number of each factor is denoted by
    a positive integer $n$, design a vector $H = ({h_1}, \cdot \; \cdot \; \cdot ,{h_m})$, where the positive integer $h_i$, $i=1,2,..., m$ is smaller than $n$, and the greatest common divisors of $n$ and $h_i$ equal to 1.
\item[(ii)] The values of the $j$th column in the uniform design table are generated by the following equation
    \begin{align}
    {u_{ij}} = (i \cdot {h_j})[\bmod \;n],
    \end{align}
    where $[\bmod \;n]$ represents congruence operation, ${u_{ij}}$ denotes the value of $ith$ rows, $jth$ columns. Specifically, $u_{ij}$ can be generated by the following recursion
    \begin{align}
    {u_{1j}} =& {h_j},\\
u_{i+1,j}=&\left\{ \begin{array}{c}
	u_{ij}+h_j,\\
	u_{ij}+h_j-n,\\
\end{array} \right. \begin{array}{c}
	u_{ij}+h_j\leqslant n\\
	u_{ij}+h_j>n,\\
\end{array}
    \end{align}	
where $i = 1,2, \cdot  \cdot  \cdot ,n - 1$ and $j = 1,2, \cdot \; \cdot \; \cdot ,m$. Then, the design of uniform design table is completed.
\end{enumerate}

\subsection{Use table of uniform design table}

In practical application, as column number $m$ is not equal to the number of variables $s$ needed to be studied, we should choose $s$ columns from $m$ columns, but there are many possibilities. Therefore, it is necessary to design a selection principle to choose some suitable data which has good evenness. Existing methods for evaluating evenness mainly include three kinds, one is based on the concept of distance, such as maximum minimum distance and minimum maximum distance, the other is based on the optimal design, such as A-optimal, D-optimal, etc., and the rest is dependent on the deviation such as $C{D_2}$, $S{D_2}$ and ${D_2}$. In this paper, the widely used centralization bias method central $C{D_2}$ bias is used to evaluate the evenness of data, and it is given as follows.
\begin{align}
CD_2=&\left[ \left( \frac{13}{12} \right) ^s-\frac{2^{1-s}}{n}\sum_{i=1}^n{\prod_{j=1}^s{\left( 2+\left| x_{ij}-\frac{1}{2} \right| \right.}} \right.
\nonumber\\
&\left. -\left( x_{ij}-\frac{1}{2} \right) ^2 \right) +\frac{1}{n^2}\sum_{i,l=1}^n{}\prod_{j=1}^s{\left( 1+\frac{1}{2}\left| x_{ij}-\frac{1}{2} \right| \right.}
\nonumber\\
&\left. \left. +\frac{1}{2}\left| x_{lj}-\frac{1}{2} \right|-\frac{1}{2}\left| x_{ij}-x_{lj} \right| \right) \right] ^{1/2},
\end{align}	
where ${x_{ij}}=\frac{{2{u_{ij}} - 1}}{{2n}}$, $i =1,2,\cdots,n$, $j = 1,2,\cdots,s$.  Accordingly, the table ${U_n}({n^s})$ with $n$ rows and $s$ columns that has the smallest deviation would be chosen.

%Let $\mathbf{x_i} = ({x_{i1}},\cdots,{x_{im}})$.
%\begin{align}
%    {x_{ij}} = &\frac{{2{u_{ij}} - 1}}{{2n}}\\
%    \mathbf{x_i} = &({x_{i1}}, \cdot \; \cdot \; \cdot ,{x_{im}})\\
%    j =& 1, \cdot \; \cdot \; \cdot ,s\nonumber\\
%    i =& 1, \cdot \; \cdot \; \cdot ,n.\nonumber
%\end{align}

% Accordingly, $n$ test points are transformed into $n$ points $\mathbf{x_1}, \cdot \; \cdot \; \cdot ,\mathbf{x_n}$ range from zero to one. Study the uniformity of the original $n$ test points is equivalent to study the uniformity of $\mathbf{x_1}, \cdot \; \cdot \; \cdot ,\mathbf{x_n}$. Finally, the data set with smallest deviation would be chosen.

%%%%%%%%%%%%%%%%%%%%%%%%%%%%%%%%%%%%%%%%%%%%%%%%%%%%%%%%%%%%%%%%%%%%%%
\section{Application of Uniform Design in Searching Controller Parameters}
In this section, we will give the detailed process of controller parameter design by using uniform design algorithm.

It should be noted that in order to determine which control parameters result in the optimal control performance, performance index should be designed. There are three commonly used performance indicators \cite{kishnani2014comparison}, and they are given by

ISE index:
 \begin{align}
 ISE = \int_0^\infty  {{e^2}} (t)dt.
 \end{align}	

IAE index:
 \begin{align}
 IAE = \int_0^\infty  {\left| {e(t)} \right|dt}.
\end{align}	

 ITAE index:
 \begin{align}
 IATE = \int_0^\infty  {t\left| {e(t)} \right|dt}.
 \end{align}
 
 \emph{Remark 1:} The performance index is only used to select the controller parameters we want most. For example, if the overshoot of the controlled system is not required, we can make the value of overshoot be the performance index. Therefore, the performance indices are not limited to the equations (5)-(7). The researcher can choose any other performance indices such as overshoot, rise time, etc. In this paper, we choose ITAE as performance index.

 The proposed search algorithm based on uniform design runs through the following steps:

\emph{Step 1:} Specify the maximum ${k_{i_{\max} }}$ and minimum ${k_{i_{\min} }}$ that each control parameter $k_i$, $i=1,2,\cdots,s$ can be chosen. Then, the level number of each parameter $n$ can be calculated from the equation $n = \frac{{{k_{i_{\max }}} - {k_{i_{\min} }}}}{\lambda_i } + 1$, where $\lambda_i $  is a positive constant and denotes the interval value between two level numbers. The value of each level can be solved from the equation ${a_{ij}} = {k_{i_{\min} }} + \lambda_i(j - 1) $, $j = 1,2, \cdots, n$. Therefore, the value of each control parameter $k_i$ can be chosen from a set $A_i=\{a_{i1},a_{i2},\cdots,a_{ij}\}$.

 \emph{Step 2:} After determining the values of $n$ and $s$ from step 1, a uniform design table ${U_n}({n^s})$ with $n$ lines and $s$ columns can be created according to the scheme described in section 2. For the purpose of convenience, a matrix $M\in R^{n\times  s}$ is used to denote the table ${U_n}({n^s})$, where $M(j,i)$ represents the number of $j$th row and $i$th column of the table.

% \emph{Step 3:} $s$ represents the variables to be discussed and thus, $s$ columns needed to be selected from $m$ columns of the uniform design table. Then, there are $C_m^s$ possibilities for possible choice. The deviation of each combination will be calculated in accordance with centralization bias method central $C{D_2}$  bias and the one with smallest bias will be choose. It should be noted that the values of this combination are not the controller parameters but their ordinal numbers.

 \emph{Step 3:} By using the matrix $M$, each control parameter is designed as $k_i=A_i(M(j,i))$, where $A_i(M(j,i))$ is the $M(j,i)$th number in set $A_i$, that is, $k_i=a_{iM(j,i)}$. Then, the response of the studied system under the control method with the designed controller parameters is obtained. Meanwhile, calculate the responding performance indices can be calculated by using equation (7). Accordingly, $n$ groups of performance indices under the $n$ groups controller parameters are obtained.

 \emph{Step 4:} Compare the values of $n$ groups of performance indices, the group of controller parameters with the smallest performance index is the optimal. That is to say, apply the optimal control parameters to the controlled system, we will obtain a optimal control performance under the designed performance index.

It can be seen that the proposed search method does not need many generations to generate new control parameters compared with the existing intelligent algorithms such as genetic algorithm and particle swarm optimization. In addition, as it has nothing to do with the form of the system models and the control methods, it is enable to design the optimal parameters of PID control method as well as parameters of other control methods.

%%%%%%%%%%%%%%%%%%%%%%%%%%%%%%%%%%%%%%%%%%%%%%%%%%%%%%%%%%%%%%%%%%%%%%
\section{Simulation Results}
To illustrate the merits of the proposed controller parameters design method more clearly, two examples are given in this section. First example is to compare the proposed method with an improved  genetic algorithm presented in \cite{srinivas1994adaptive}. Second example is to illustrate the effectiveness of the proposed controller design method for the controlled system which is nonlinear, strongly coupled and multivariable.

\emph{Example 1:} In this example, a 3-DOF helicopter simulation platform is used, and the dynamics of the elevation and pitch axis are given as follows
\begin{align}
{J_e}\ddot \alpha  &= {K_f}{l_a}\cos (\beta )({V_f} + {V_b}) - mg{l_a}\cos (\alpha )\nonumber\\
& = {K_f}{l_a}\cos (\beta ){V_1} - mg{l_a}\cos (\alpha ),\\
{J_p}\ddot \beta  &= {K_f}{l_h}({V_f} - {V_b})\nonumber\\
 &= {K_f}{l_h}{V_2},
\end{align}
where $\alpha $ and $\beta $ are elevation and pitch angle, respectively. ${J_e}$ and ${J_p}$ denote the inertial of the elevation and pitch axis, ${K_f}$ is the force constant produced by motor, ${l_a}$ presents the distance from elevation axis to the center of 3-DOF body and ${l_h}$ is the distance from the pitch axis to either motor, ${V_f}$ and ${V_b}$ are the voltages applied to the front and back motors and ${V_1} = {V_f} + {V_b}$ represents the sum of voltages, ${V_2}={V_f} - {V_b}$ is the difference between the two voltages, $m$ is the effective mass of the 3-DOF helicopter and $g$ is the gravitational acceleration constant. The pitch angle is limited to $( - \pi /2,\pi /2)$ mechanically.

\emph{Remark 2:} The studied system has two inputs and three outputs. The inputs are the voltages of two spiral motors, and the outputs are three attitude angles (pitch angle, elevation angle and travel angle). In this paper, the main purpose is to study the proposed search algorithm and thus, elevation and pitch axis are investigated while the travel axis is set to free.

For the purpose of simplicity, define ${x_1} = \alpha $, ${x_2} = \dot \alpha $, ${x_3} = \beta $, ${x_4} = \ddot \beta $, ${u_1} = {V_1}$ and ${u_2} = {V_2}$, elevation and pitch dynamics can be rewritten as the following form
\begin{align}
{{\dot x}_1} =& {x_2}\nonumber\\
{{\dot x}_2} =& {K_f}{l_a}\cos ({x_3}){u_1} - mg{l_a}\cos ({x_1})\nonumber\\
{{\dot x}_3} =& {x_4}\nonumber\\
{{\dot x}_4} =& {K_f}{l_h}{u_2}.
\end{align}

Choose the elevation and pitch angle references as ${r_{ele}}=0.4$ and ${r_{pit}}=0.02$, the tracking errors are defined as ${e_{ele}}={r_{ele}}-x_1$ and ${e_{pit}}={r_{pit}}-x_3$, respectively. The controllers are designed by using the wildly used PID algorithm, and they are given by
\begin{align}
u_1 =& {k_{{p_{ele}}}}{e_{ele}} + {k_{{d_{ele}}}}\frac{{d{e_{ele}}}}{{dt}} + {k_{{i_{ele}}}}\int {{e_{ele}}} dt\\
u_2 =& {k_{{p_{pit}}}}{e_{pit}} + {k_{{d_{pit}}}}\frac{{d{e_{pit}}}}{{dt}} + {k_{{i_{pit}}}}\int {{e_{pit}}} dt,
\end{align}	
where ${k_{{p_{ele}}}}$, ${k_{{d_{ele}}}}$, ${k_{{i_{ele}}}}$, ${k_{{p_{pit}}}}$, ${k_{{d_{pit}}}}$ and ${k_{{i_{pit}}}}$ are positive constants and represent proportional, integral and derivative gains. It should be noted that each control parameter is regarded as a factor, the value of each parameter is called the level of this parameter.

\begin{table}[ptb]
\caption{System parameters of 3-DOF helicopter \label{tab1}}
{\begin{tabular*}{\linewidth}{@{\extracolsep{\fill}}lll@{}}\toprule
Parameter &Value \\
\midrule
$k_f$     & $0.1188N/V$              \\
$L_a$     & $0.660m$                     \\
$J_e$     &  $1.034kg \cdot {m^2}$   \\
$J_p$     &  $0.045kg \cdot {m^2} $                    \\
$g$       &  $9.8m/s$                    \\
$m$       &  $0.094kg$                   \\
$L_h$     &  $0.178m$                       \\
%$m$ & $0.094$ kg\\\hline\hline
\midrule
\end{tabular*}}{}
\end{table}

\begin{table}[htbp]
\caption{Range of control parameters\label{tab2}}
{\begin{tabular*}{\linewidth}{@{\extracolsep{\fill}}lll@{}}\toprule
Control Parameter &Min Value & Max Value\\
\midrule
${k_{ele\_p}}$     & $0$   &$60$           \\
${k_{ele\_d}}$     & $0$   &$60$                   \\
${k_{ele\_i}}$     & $0$   &$6$     \\
${k_{pit\_p}}$     & $0$   &$60$                \\
${k_{pit\_d}}$     & $0$   &$60$               \\
${k_{pit\_i}}$     & $0$   &$30$             \\
\midrule
\end{tabular*}}{}
\end{table}

\begin{table}[htbp]
\caption{Search results of proposed method\label{tab3}}
{\begin{tabular*}{\linewidth}{@{\extracolsep{\fill}}lll@{}}\toprule
Control Parameter &Case 1 & Case 2\\
\midrule
${k_{ele\_p}}$     & $55.8$   &$58.1$           \\
${k_{ele\_d}}$     & $37.8$   &$43.5$                   \\
${k_{ele\_i}}$     & $22.5$   &$18.35$     \\
${k_{pit\_p}}$     & $38.8$   &$48.9$                \\
${k_{pit\_d}}$     & $14.4$   &$15.3$               \\
${k_{pit\_i}}$     & $0.00$   &$0.13$             \\
$n$                & $301$    & $601$ \\
${\sigma _{ele}}$  & $4.51\%$ & $1.46\%$  \\
${\sigma _{pit}}$  & $0.84\%$   & $0.00\%$ \\
${e_{ele}}$        & $38.75$  & $33.87$ \\
${e_{pit}}$        & $0.84$   & $0.87$ \\
$t$                & $4.6347$ & $9.99$\\
\midrule
\end{tabular*}}{}
\end{table}

\begin{table*}[ptb]
\caption{Controller parameters.}\label{tab4}
\centering
\begin{tabular*}{\textwidth}{@{}@{\extracolsep{\fill}}cccccccccccccc@{}}
\toprule
Control Parameter &1  &2  &3  &4  &5  &6\\
\midrule
${k_{ele\_p}}$     & $47.38$   &$44.23$      &$44.52$    &$56.59$    &$42.42$     &$52.49$ \\
${k_{ele\_d}}$     & $30.88$   &$40.46$      &$35.82$    &$44.86$    &$29.66$     &$39.40$ \\
${k_{ele\_i}}$     & $19.61$   &$13.98$     &$19.24$     &$19.84$    &$17.09$     &$18.22$ \\
${k_{pit\_p}}$     & $58.64$   &$58.03$      &$33.51$    &$42.66$    &$55.84$     &$57.66$ \\
${k_{pit\_d}}$     & $20.29$   &$23.08$      &$10.24$    &$16.02$    &$9.99$      &$23.95$ \\
${k_{pit\_i}}$     & $0.03$   &$0.69$       &$0.01$     &$0.19$      &$0.59$      &$0.29$  \\
${\sigma _{ele}}$  & $0.00\%$ &$0.23\%$    &$6.64\%$      &$2.64\%$  &$3.31\%$    &$0.10\%$\\
${\sigma _{pit}}$  & $5.01\%$  &$0.66\%$   &$4.55\%$      &$0.46\%$  &$10.03\%$   &$1.02\%$ \\
${e_{ele}}$        & $51.09$  & $62.95$    &$67.16$      &$49.06$    &$54.56$     &$36.43$ \\
${e_{pit}}$        & $0.79$   & $1.72$     &$0.90$      &$1.14$      &$1.11$      &$1.35$  \\
$t$                & $158.00$ & $164.85$   &$163.33$      &$165.26$  &$159.92$    &$163.43$ \\
\midrule
\end{tabular*}
\end{table*}

\begin{table}[ptb]
\caption{Controller parameters of UAV \label{tab1}}
{\begin{tabular*}{\linewidth}{@{\extracolsep{\fill}}cccccccccccccc@{}}\toprule
Parameter &Value & Parameter & Value \\
\midrule
$k_1$     & $7.0$  & $z_{1}$      &$2.1285$     \\
$k_2$     & $7.0$     &   $z_{3}$     &  $2.0952$      \\
$k_3$     &  $7.0$ & $z_{5}$ & $0.0650$\\
$k_4$     &  $7.0$  & $z_{7}$ & $ 1.0343$               \\
$k_5$       &  $7.0$    &$z_{9}$&    $1.1449$            \\
$k_6$       &  $7.0$    &$z_{11}$&     $0.9980$          \\
$k_7$     &  $7.0$        &$$&      $$         \\
$k_8$     &  $24.0$        &$$&     $$          \\
$k_9$     &  $20.8$        &$$&     $$          \\
$k_{10}$     &  $17.6$      &$$&      $$           \\
$k_{11}$     &  $24.0$      &$$&        $$         \\
$k_{12}$     &  $12.8$       &$$&       $$         \\
%$m$ & $0.094$ kg\\\hline\hline
\midrule
\end{tabular*}}{}
\end{table}

Generally, they are designed by using trial and error, which is quite time-consuming and experience dependent. However, they can be easily determined by using the proposed method. Specifically, the range of each control parameter is shown in Table~\ref{tab2}. In order to compare the effects of different experiment times $n$ on the control performance, two cases are given. In case 1, the experiment times $n_1=301$, while in another case, the experiment times $n_2=601$. When choose IATE criterion (see equation (7)) as the performance index, the optimal controller parameters under this performance index of these two cases can be searched by using the steps described in Section 3. The research results are shown in Table 3. ${\sigma _{ele}}$ and ${\sigma _{pit}}$ represent overshoot, $e_{ele}$ and $e_{pit}$ denote the performance index calculated by using equation (7), while $t$ is the time taken to search for the optimal result. The tracking performances of elevation and pitch angle by using PID controller with the controller parameters shown in Table 3 are shown in Figs. 1 and 2.

Moreover, in order to illustrate the effectiveness of the proposed method, an improved GA, which can overcome the local optimum of traditional GA, proposed in paper \cite{srinivas1994adaptive} is used as a comparative method. In such case, the range of each parameter is also chosen as what is shown in Table 2. Since each generation is randomly generated, the optimal result obtained by GA of each experiment is different. Hence, six experiments (see Table~\ref{tab4}) were conducted to ensure the reliability of the comparison between GA and the proposed method. The tracking performances of elevation and pitch angle by using PID controller with the controller parameters shown in Table 4 are shown in Figs. 1 and 2.

 \begin{figure}[htbp]
\centering
{\includegraphics[width=7.5cm,height=6cm]{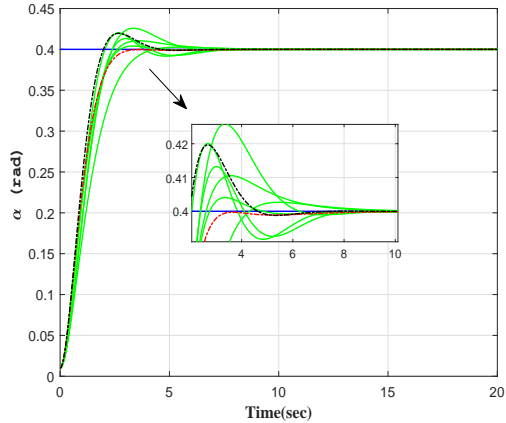}
\caption{ Tracking performance of elevation angle by using PID control method with the controller parameters searched by proposed method (black dashed line and red dashed line represent control performance by using the controller parameters in case 1 and case 2, respectively) and improved GA method in \cite{srinivas1994adaptive} (six green dashed lines denote the control performance by using the controller parameters shown in Table 4).\label{fig3}}}
\centering
\end{figure}

\begin{figure}[htbp]
\centering
\includegraphics[width=7.5cm,height=6cm]{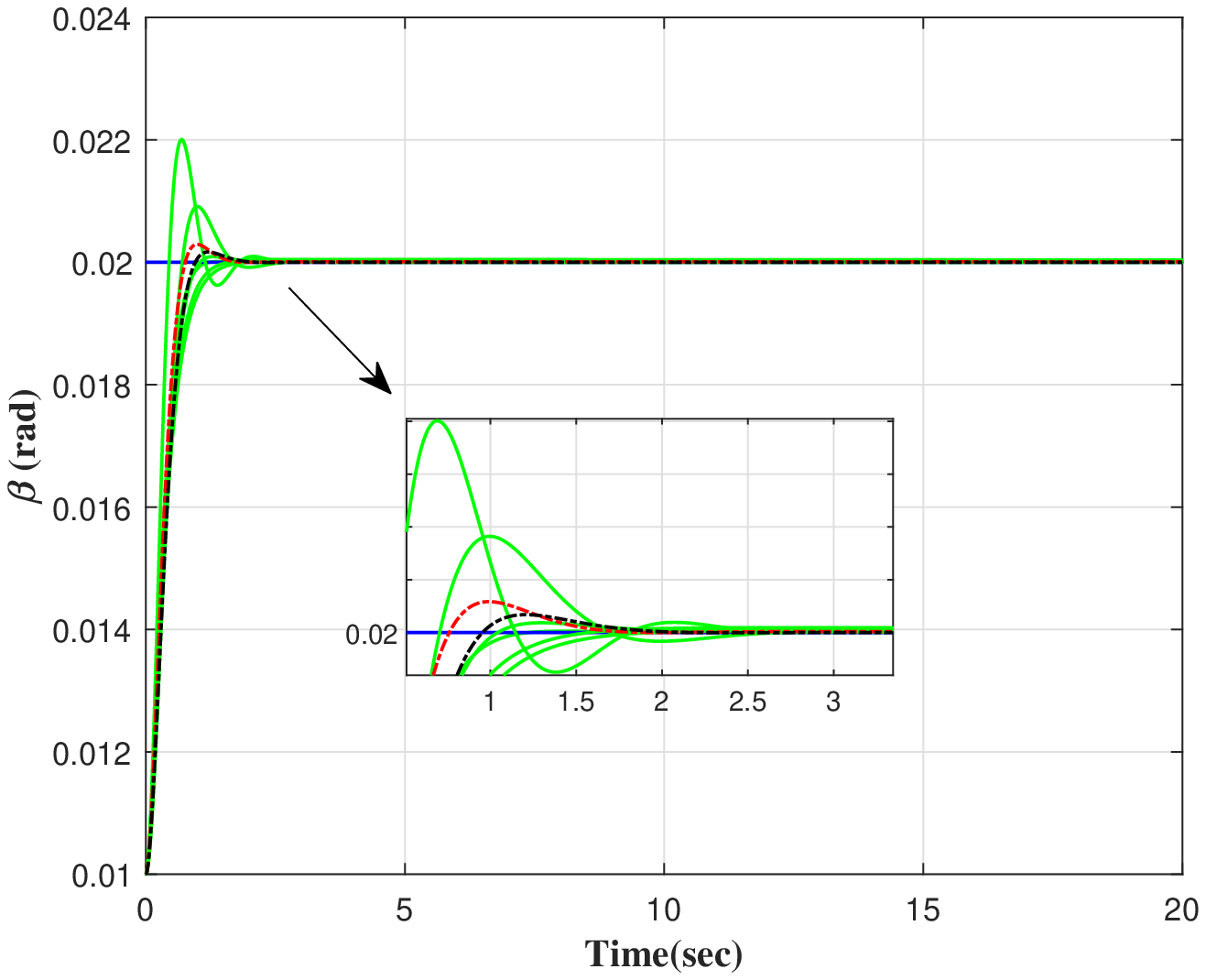}\caption{ Tracking performance of pitch angle by using PID control method with the controller parameters searched by proposed method (black dashed line and red dashed line represent control performance by using the controller parameters in case 1 and case 2, respectively) and improved GA method in \cite{srinivas1994adaptive} (six green dashed lines denote the control performance by using the controller parameters shown in Table 4).\label{fig4}}%%
\end{figure}

\begin{figure}[htbp]
\centering
\includegraphics[width=7.5cm,height=6cm]{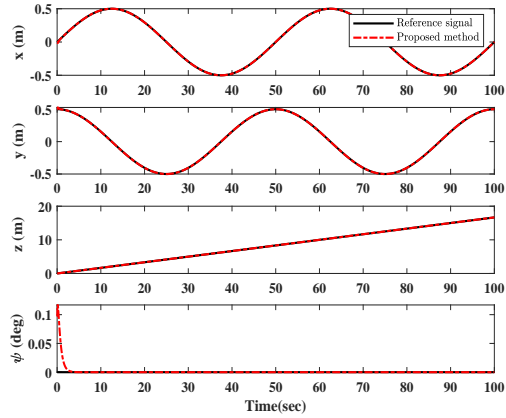}\caption{ Tracking performance by using back-stepping control method with the controller parameters searched by proposed method.\label{fig4}}%%
\end{figure}

\begin{figure}[htbp]
\centering
\includegraphics[width=7.5cm,height=6cm]{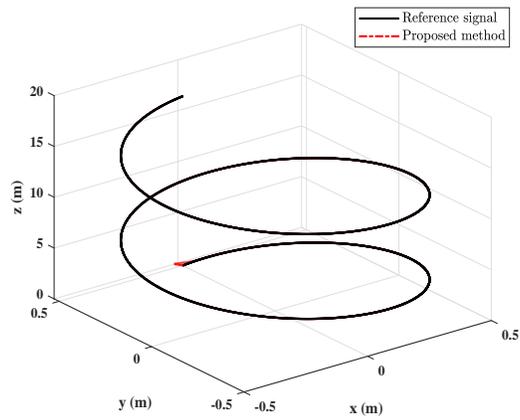}\caption{ Space diagram of position by using back-stepping control method with the controller parameters searched by proposed method.\label{fig4}}%%
\end{figure}

\begin{figure}[htbp]
\centering
\includegraphics[width=7.5cm,height=6cm]{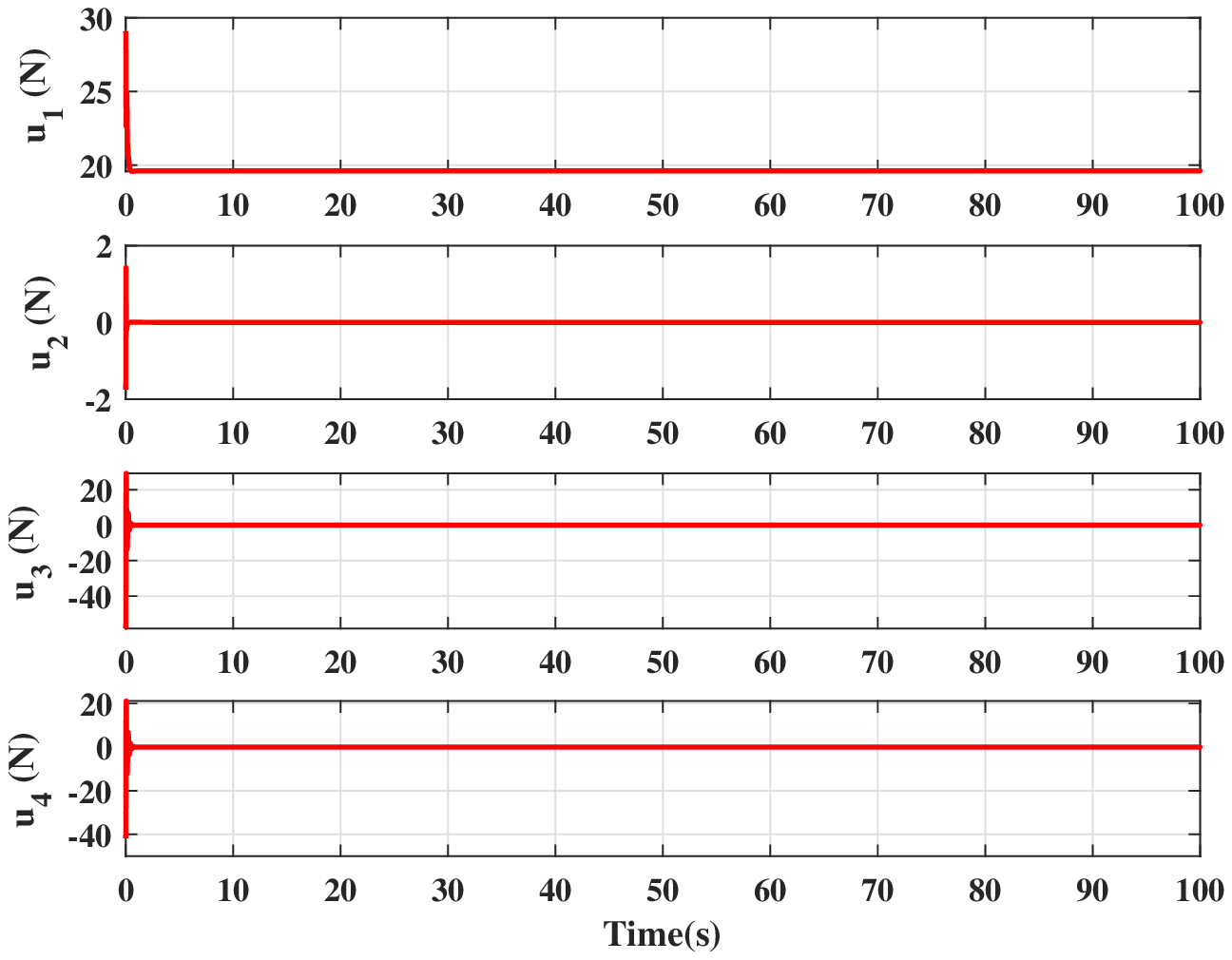}\caption{ Control inputs by using Back-stepping control method with the controller parameters searched by proposed method.\label{fig4}}%%
\end{figure}

It can be found from Table 3 that compared with the case when $n=301$, the case when $n=601$ has a better control performance for the overshoots and tracking errors are smaller. But it at the expense of more time, which needs $9.99s$ to search the optimal controller parameters. However, case 1 only needs $4.6347s$. Table 4 shows that the controller parameters obtained from each experiment is different. In some cases, the values of overshoots or tracking errors are similar as that by using proposed method, but most of them have a worse control performance. Furthermore, search controller parameter by using this method needs about $160s$, which is quite longer nearly $16$ times more than our proposed method (one needs $4.6347s$, another needs $9.99s$). Figs. 1 and 2 clearly compare the control performance by using PID controller with different controller parameters. In conclusion, the proposed method is more effective, easier to use and time saving.

\emph{Example 2:} In this example, a 6-DOF quadrotor Unmanned Aerial Vehicle (UAV) system is used for its model is nonlinear, strongly coupled, multivariable and under-actuator. The object of this example is to design a set of controller parameters by using proposed method such that a satisfactory output tracking performance can be obtained via back-stepping control method.

The dynamic model of the quadrotor UAV system is given by
\begin{align}
\begin{cases}
	\ddot{\varphi}=\dot \vartheta \dot\psi \left( \frac{I_y-I_z}{I_x} \right) +\frac{l}{I_x}U_2\\
	\ddot{\vartheta}=\dot\varphi \dot\psi \left( \frac{I_z-I_x}{I_y} \right) +\frac{l}{I_x}U_3\\
	\ddot{\psi}=\dot\vartheta \dot\varphi \left( \frac{I_x-I_y}{I_z} \right) +\frac{l}{I_x}U_4\\
	\ddot{x}=U_1\left( \cos \psi \sin \vartheta \cos \varphi +\sin \psi \sin \varphi \right) /m\\
	\ddot{y}=U_1\left( \sin \psi \sin \vartheta \cos \varphi -\cos \psi \sin \varphi \right) /m\\
	\ddot{z}=U_1\left( \cos \psi \cos \varphi \right) /m-g,\\
\end{cases}
\end{align}
where $\varphi$, $\vartheta$ and $\psi$ are pitch, roll and yaw angles, $x$, $y$ and $z$ represent the position of the quadrotor UAV; $g$ is the gravity constant; $l$ represents the distance from the center of mass to each motor; $m$ is the mass of the quadrotor UAV body; $I_x$, $I_y$ and $I_z$ are the moment of inertial with respect to each axis, the system parameters are given in Table 1.

In order to overcome the obstacles caused by underactuation problem, we define the following intermediate control laws
\begin{align}
U_x=&U_1\left( \cos \psi \sin \vartheta \cos \varphi +\sin \psi \sin \varphi \right)\\
U_y=&U_1\left( \sin \psi \sin \vartheta \cos \varphi -\cos \psi \sin \varphi \right)\\
U_z=&U_1\left( \cos \psi \cos \varphi \right).
\end{align}

Therefore, we obtain the controller $U_1$ as follows
\begin{align}
U_1=\sqrt{U_{x}^{2}+U_{y}^{2}+U_{z}^{2}}.
\end{align}

Calculate the desired roll and pitch angles as
\begin{align}
\varphi _d=&\sin ^{-1}\left( \frac{1}{U_1}\left( U_x\sin \left( \psi _d \right) -U_y\cos \left( \psi _d \right) \right) \right)
\\
\vartheta _d=&\tan ^{-1}\left( \frac{1}{U_z}\left( U_x\cos \left( \psi _d \right) +U_y\sin \left( \psi _d \right) \right) \right).
\end{align}

For the purpose of convenience, let
\begin{align}
\bar \chi=&[\chi_1, \chi_2, \chi_3, \chi_4, \chi_5, \chi_6, \chi_7, \chi_8, \chi_9, \chi_{10}, \chi_{11}, \chi_{12}]^T\nonumber\\
=&[\varphi, \dot \varphi, \vartheta, \dot \vartheta, \psi, \dot \psi, x, \dot \psi, y, \dot y, z, \dot z]^T\nonumber
\end{align}
be the system state vector. Choose the desired references as $\chi_{1d}=\varphi _d$, $\chi_{2d}=\vartheta _d$, $\chi_{5d}=0$, $\chi_{7d}=0.5sin(\pi/25 t)$, $\chi_{9d}=0.5cos(\pi/25 t)$ and $\chi_{11d}=1/6 t$. Define the tracking error as
\begin{align}
z_{i}=&\chi_i-\chi_{id},  i=1,2,\cdots,12.\nonumber
%z_{i,2}=&\chi_i-\alpha_{i/2}, i=2,4,\cdots,12\nonumber
\end{align}

By using back-stepping control method, the controller can be designed as follows
\begin{align}
\chi_{id}=&-k_{i-1}z_{i-1}+ \dot \chi_{(i-1)d},  i=2,4,\cdots,12,\\
U_{i/2}=&\frac{1}{g_{i/2}(\bar \chi)}(-k_{i}z_i-f_{i/2}(\bar \chi)+\dot \chi_{(i)d}-z_{i-1}).
\end{align}

Let IATE criterion be the performance index, the optimal controller parameters searched by using the proposed method are shown in Table 5, the tracking performance by using the back-stepping control method with the optimal controller parameters are shown in Figs. 3 and 4. While, Fig.5 shows the control input. It can be found from the simulation results that the controller designed as equations (20) and (21) with the designed parameters can results in a quite satisfactory output tracking performance.

%
%Design the updating laws of FNN and controllers by using the proposed method in this paper, and the control parameters are chosen as $k_{x_1}=7$, $k_{x_2}=7$, $k_{y_1}=7$, $k_{y_2}=7$, $k_{z_1}=7$, $k_{z_2}=7$, $k_{psi_1}=7$, $k_{psi_2}=24$, $k_{\vartheta_1}=20.8$, $k_{\vartheta_2}=17.6$, $k_{\varphi_1}=24$, $k_{\varphi_2}=12.8$. The simulation results are given in Figs. 1-3. Meanwhile, the controller designed by busing an ANN-based method is designed and the simulation results are also shown in the figures for comparison. It can be found from Fig. 1 that the proposed has a better tracking performance than the ANN-based method for the reason that it is more excellent in approximating unknown dynamics such that the unknown information can be better compensated.

%%%%%%%%%%%%%%%%%%%%%%%%%%%%%%%%%%%%%%%%%%%%%%%%%%%%%%%%%%%%%%%%%%%%%%
\section{Conclusion}
%\section{Conclusion}

%Especially when this part is nested in the system, it can greatly short the whole process time from controller parameter selection to system stability and help achieve finite time control.

Simulation results show that the proposed novel controller parameter research method via UD algorithm is quite effective. Its superiorities are listed as follows: 1) It does not require multiple generations, and can find the optimal controller parameters one time, which is very efficient and time saving. 2) Due to the introduction of uniform design method, the experiment times can be largely reduced compared with exhaustive method. 3) It is easy to operate, and does not require profound theoretical knowledge once the uniform design table is designed.  4) Since this method is an optimization method which has nothing to do with the control method and the model of the controlled plants, it can be generalized to a wider range of control methods and systems.

%However, as one kind of research algorithms, it has disadvantages that it aims to find the most comprehensive combinations with the least data and thus, some combinations of control parameters may not be taken into consideration when there are less samples in the range of controller gains so that it cannot exclude the possibility of better results. This problem can be overcomed by setting small parameter intervals, but it is at the expense of increased computation (see Table~\ref{tab3}).
%
%Therefore, one of the next efforts is to improve the proposed tuning control parameter based on uniform design. The other is that apply this method into practical system to test its performance. Due to the uncertainty and inaccuracy of modeling, the actual system and the simulation system are not completely consistent, which results in the optimal controller parameters obtained by simulation may not suitable for the actual system. It is impractical to put parameters directly into the actual system to calculate their fitness and find optimal gains, so a threshold can be set in practical application. When the amplitude of the state variable is greater than a certain value, the program terminates to prevent equipment damage. Therefore, this method will be further studied and verified in the actual system.

\section*{References}

\bibliography{reference}

\end{document}